\documentstyle[aps,epsfig,floats]{revtex}

\begin{document}
\draft
\title{
Consistency of spectroscopic factors from (e,e$'$p) reactions \\
at different momentum transfers
}

\author{Marco Radici}
\address{Dipartimento di Fisica Nucleare e Teorica, 
Universit\`{a} di Pavia, and\\
Istituto Nazionale di Fisica Nucleare, 
Sezione di Pavia, I-27100 Pavia, Italy}

\author{W.H. Dickhoff}
\address{Department of Physics, Washington University, St. Louis, Missouri 63130, USA}

\author{E. Roth Stoddard}
\address{Department of Physics, University of Missouri at Kansas City, 
Kansas City, Missouri 64110, USA}

\date{\today}

\maketitle

\begin{abstract}

The possibility to extract relevant information
on spectroscopic factors from (e,e$'$p) reactions at high $Q^2$ is
studied. Recent ${}^{16}$O(e,e$'$p) data at $Q^2 = 0.8$ (GeV/$c)^2$
are compared to a theoretical approach which includes an eikonal
description of the final-state interaction of the proton, a microscopic
nuclear matter calculation of the damping of this proton, and 
high-quality quasihole wave functions for $p$-shell nucleons in ${}^{16}
{\rm O}$.
Good agreement with the $Q^2 = 0.8$ (GeV/$c)^2$ data is obtained when
spectroscopic factors are employed which are identical to those required to 
describe earlier low $Q^2$ experiments.

\end{abstract}


\section{Introduction}
\label{sec:intro}

During the last fifteen years a quiet revolution has taken place in the
perception of the nucleus. During this period the study of the (e,e$'$p) reaction has
clarified the limits of validity of the mean-field description of nuclei.
In particular, absolute spectroscopic factors have been obtained for the
removal of protons from many nuclei~\cite{diehu,sihu,lap,pasihu}.
The qualitative features of the strength distribution suggest a considerable 
mixing between single-hole states and more complicated configurations, like
two-hole$-$one-particle states.
The resulting fragmentation pattern of the single-particle (sp) strength
exhibits a single peak carrying about 65-70\% of the strength for states
in the immediate vicinity of the Fermi energy, while more deeply bound orbitals
display a strongly fragmented distribution reminiscent of a complex sp energy.
In addition to these fragmentation features, an important depletion of the sp
strength has been established which is associated with ground-state correlations
induced by strong short-range and tensor correlations~\cite{von}. 
This leads to an overall reduction of the sp strength for all mean-field
orbits in all nuclei by 10-15\%~\cite{dimu,dic1}.
This theoretical result has recently been confirmed also for deeply-bound
orbits in an (e,e$'$p) experiment on ${}^{208}{\rm Pb}$ in a wide domain of
missing energy and momentum~\cite{lapx,bat}.

The analysis of the (e,e$'$p) data has relied on the Distorted Wave Impulse
Approximation (DWIA) for both the Coulomb distortion of the electron waves
(in heavy nuclei) and the outgoing proton~\cite{bgprep,book,kelly,frumou}.
The proton distortion is described in terms of an optical potential required
to describe elastic proton scattering data at relevant energies~\cite{lap}.
There is some uncertainty related to this treatment since elastic proton
scattering is considered to be a surface reaction and no detailed information
is obtained related to the interior of the nucleus.
This uncertainty gives rise to an estimated error of about 10\%.
Such an estimate may be inferred by considering the difference between
the relativistic and nonrelativistic treatment of the proton distortion.
It is shown in Ref.~\cite{udias} that this difference is essentially
due to the reduction of the interior wave function in the relativistic
case. This feature can also be generated by including a reasonable amount of
nonlocality in the optical potential~\cite{udias}.

A serious challenge to the interpretation of (e,e$'$p) experiments
was recently published in Refs.~\cite{lapy,fsz}.
This challenge consists in questioning the validity of the constancy of the
spectroscopic factor as a function of the four-momentum (squared), $Q^2$,
transferred by the virtual photon to the knocked-out nucleon.
In Ref.~\cite{lapy} a conventional analysis of the world's data for (e,e$'$p)
experiments on ${}^{12}{\rm C}$ at low $Q^2$ generated results 
consistent with previous expectations.
Data at higher values of $Q^2$ were then analyzed within the framework
of a theoretical model which employs Skyrme-Hartree-Fock bound-state
wave functions for the initial proton, a Glauber-type description of the
final-state interaction of the outgoing proton, and a factorization
approximation for the electromagnetic vertex~\cite{lapy,fsz}.
Within the framework of this theoretical description, spectroscopic factors
were obtained which increase substantially with increasing $Q^2$ for the
${}^{12}{\rm C}$ nucleus.

The spectroscopic factor is a many-body
quantity defined without reference to a probe.
For valence hole states in nuclei it simply represents the probability for the
removal of a nucleon with prescribed
quantum numbers from the ground state of the target while ending up
in a state of the nucleus with one particle less.
In the conventional analysis of the experimental data these quantum numbers
involve the trivial values of parity and total angular momentum but also
require the corresponding wave function to be a solution of a Woods-Saxon 
potential at the appropriate binding energy.
This potential is adjusted to generate an optimum description of the
shape of the experimental cross section.
The resulting theoretical representation of the cross
section must then be multiplied by a constant factor to coincide
with the experimental cross section.
This constant factor is then interpreted as the spectroscopic factor.
Another important ingredient in this analysis is the choice of the 
electron-proton cross section which must be considered off-shell~\cite{defor}.
This leads to a small additional uncertainty in the analysis of low $Q^2$ data
as discussed recently in Ref.~\cite{udiasa}.

Further clarification of this intriguing situation with different 
spectroscopic factors at different $Q^2$ is urgently needed.
For this purpose we consider in this paper a study of recently
published (e,e$'$p) data for ${}^{16}{\rm O}$ at $Q^2 = 0.8$ 
(GeV/$c)^2$~\cite{e89003}. 
An unfactorized approach is used, as it is required at low 
$Q^2$~\cite{book} and has been recently advocated also for high $Q^2$
reactions in Ref.~\cite{jes}. However, the higher energy 
of the outgoing proton requires 
a description which contains different elements than the conventional 
low $Q^2$ analysis. 
For the electromagnetic current operator we follow 
the approach of Ref.~\cite{kelly1}, where a relativistic current 
operator was used in a Schr\"odinger-based calculation, avoiding any 
nonrelativistic reduction and including the effect of spinor distortion 
by the Dirac scalar and vector potentials (see also 
Refs.~\cite{heda-poor,vanord,bgp,jin,udias}). As for the final state, 
we employ a recently developed eikonal description of the final-state
interaction (FSI) of the proton with the nucleus that has been tested 
against DWIA solutions of a complex spin-dependent optical 
potential~\cite{br3,br2}. 
The absorption of the proton is described theoretically by linking it to
the corresponding absorption of a nucleon propagating through nuclear
matter. The relevant quantity is the nucleon self-energy which is obtained from
a self-consistent calculation of nucleon spectral functions including the
effects of realistic short-range and tensor correlations~\cite{libth}.
This description of FSI is combined with previous results for the bound-state
wave functions of the $p$-shell quasihole states in
${}^{16}{\rm O}$~\cite{mpd}, that have been deduced by solving the Dyson 
equation with a nucleon self-energy containing the same short-range correlations  
but in a finite volume.  
In Ref.~\cite{previous} these wave functions have been used to analyze low $Q^2$
data for the ${}^{16}$O(e,e$'$p) reaction~\cite{nikhef}. In the present work, 
these very same wave functions produce a good description of the 
shape of the coincidence cross section for the $p$-shell quasihole states 
at high $Q^2$~\cite{e89003} using the same spectroscopic factors obtained from 
the low $Q^2$ data~\cite{nikhef}. A preliminary result was reported in 
Ref.~\cite{lund}.

A consistent analysis of low and high $Q^2$ data requires an approach in which
the same quasihole wave functions and corresponding spectroscopic factors
are used in both cases. 
In this paper we present such an approach.
In Sec.~\ref{sec:model} we discuss the ingredients of the theoretical 
description including the structure of the electromagnetic current 
(Sec.~\ref{sec:current}), the eikonal approximation (Sec.~\ref{sec:eikonal}), and 
the final hadronic tensor (Sec.~\ref{sec:tensor}).
The many-particle ingredients are discussed in Sec.~\ref{sec:spectral}
which includes a summary of the calculation of the quasihole wave functions
in Sec.~\ref{sec:quasihole}, a description of the construction of the
nonlocality factor required by a treatment of relativistic effects in
Sec~\ref{sec:dr}, and, finally, an overview of the ingredients to describe
the damping of high-momentum protons in nuclear matter 
(Sec.~\ref{sec:damping}). 
The results are discussed in Sec.~\ref{sec:results} while final conclusions
are drawn in Sec.~\ref{sec:end}.

\section{The model}
\label{sec:model}

In the one-photon exchange approximation, the differential cross
section for the scattering of an ultrarelativistic electron with 
initial (final) momentum $\vec p_e (\vec p_e^{\, \prime})$, off a nuclear 
target from which a nucleon is ejected with final momentum 
$\vec p_N^{\, \prime}$, reads~\cite{bgprep,book,kelly}
\begin{equation}
{ {d \sigma} \over {d \vec p_e^{\, \prime} d \vec p_N^{\, \prime} } }= 
{ e^4 \over {16 \pi^2}} {1 \over Q^4 p_e p'_e } \ \  
\lower7pt\hbox{$_{\lambda,\lambda'=0,\pm 1}$} \kern-28pt 
{\hbox{\raise2.5pt \hbox{$\sum$}}} 
\qquad  L_{\lambda,\lambda'} W_{\lambda,\lambda'} \  , 
\label{eq:cross}
\end{equation}
where $Q^2 = q^2 - \omega^2$ and $\vec q = \vec p_e - \vec p_e^{\, \prime}, 
\  \omega = p^{}_e - p'_e$ are the momentum and energy
transferred to the target nucleus, respectively. The lepton tensor 
$L_{\lambda,\lambda'}$ and hadron tensor $W_{\lambda,\lambda'}$ 
are conveniently expressed in the basis of unit vectors
\begin{equation}
e_0 = \left( 1, 0, 0, 0 \right) \  , \quad 
e_{\pm 1} = \left( 0, \mp \sqrt{\textstyle{1\over 2}}, -
\sqrt{\textstyle{1\over 2}} i, 0 \right) \  ,
\label{eq:basis}
\end{equation}
which define the longitudinal (0) and transverse $(\pm 1)$
components of the nuclear response with respect to the
polarization of the exchanged virtual photon. The hadron tensor 
is defined as~\cite{bgprep,book,frumou}
\begin{equation}
W^{}_{\lambda,\lambda'} = \  (-)^{\lambda + \lambda'} \  
e^\mu_\lambda e^{\nu \, *}_{\lambda'} \quad   
{\lower7pt\hbox{$_i$}} \kern-7pt {\hbox{\raise7.5pt
\hbox{$\overline \sum$}}} \
\hbox {\hbox {$\sum$} \kern-15pt
{$\displaystyle \int_f$\ } }
J^{}_{\mu}  J^*_{\nu}  \  \delta
\left( E^{}_i - E^{}_f \right) \   ,
\label{eq:hadtens}
\end{equation}
i.e. it involves the average over initial states and the sum over
the final undetected states (compatible with energy-momentum
conservation) of bilinear products of the scattering amplitude
$J^\mu$.

This basic ingredient of the calculation is built from 
the matrix element of the nuclear charge-current density operator 
$\hat J^\mu$ between the initial, $\vert \Psi^A_0 \rangle$, 
and the final, $\vert \Psi^A_f \rangle$, nuclear 
states. This complicated $A$-body problem can be simplified by 
projecting out of the Hilbert space the specific channel that 
corresponds to the experimental asymptotic conditions of a 
knocked-out nucleon with momentum $\vec p_N^{\, \prime}$ and of a residual 
nucleus, recoiling with momentum $-\vec p_m = \vec q - \vec p_N^{\, \prime}$ 
and mass $M_R$, in a well-defined state $\vert \Psi^{A - 1}_n (E_R) \rangle$ 
with energy $E_R$ and quantum numbers $n$. The scattering 
amplitude can be rewritten in a one-body representation (in momentum 
space and omitting spin degrees of freedom for simplicity) 
as~\cite{bccgp,book}
\begin{equation}
J^{\mu}_n (\omega ,\vec q, \vec p_N^{\, \prime}, E_R) = \int d \vec p 
\  d \vec p^{\, \prime} \  
\chi^{\left( -\right)\, *}_{p'_{\scriptscriptstyle N} 
E_{\scriptscriptstyle R} n} (\vec p^{\, \prime}) \  {\hat J}_{\rm eff}^{\mu}
(\vec p, \vec p^{\, \prime}, \vec q, \omega) \  
\phi^{}_{E_{\scriptscriptstyle R} n} (\vec p) \  [S_n(E_R)]^{1\over 2} \  , 
\label{eq:scattampl1}
\end{equation}
provided that ${\hat J}^{\mu}$ is substituted by an appropriate
effective one-body charge-current density operator
${\hat J}_{\rm eff}^{\mu}$, which guarantees the
orthogonality between $\vert \Psi^A_0 \rangle$ and
$\vert \Psi^A_f \rangle$ besides taking into account 
effects due to truncation of the Hilbert space. Usually, the 
orthogonality defect is negligible in standard kinematics 
for (e,e$'$p) reactions~\cite{bccgp,bgprep,book}; in any case, 
${\hat J}_{\rm eff}^{\mu}$ is 
here approximated by a one-body relativistic current operator 
including spinor distortion along the lines described in the 
following Sec.~\ref{sec:current}. 

The functions
\begin{eqnarray}
[S_n (E_R)]^{1\over 2} \phi^{}_{E_{\scriptscriptstyle R} n} 
(\vec p) &= &\langle \Psi^{A - 1}_n (E_R) \vert a (\vec p) 
\vert \Psi^A_0 \rangle \  , \nonumber \\
\chi^{\left( - \right)}_{p'_{\scriptscriptstyle N} 
E_{\scriptscriptstyle R} n} (\vec p) &= &\langle \Psi^{A - 1}_n 
(E_R) \vert a (\vec p) \vert \Psi^A_f \rangle \  ,
\label{eq:specampl}
\end{eqnarray}
describe the overlap between the residual state 
$\vert \Psi^{A - 1}_n (E_R) \rangle$ and the hole produced in 
$\vert \Psi^A_0 \rangle$ and $\vert \Psi^A_f \rangle$, 
respectively, by removing a particle with momentum $\vec p$. 
Both $\phi^{}_{E_{\scriptscriptstyle R} n}$ and 
$\chi^{\left( - \right)}_{p'_{\scriptscriptstyle N} 
E_{\scriptscriptstyle R} n}$ are eigenfunctions of a 
Feshbach-like nonlocal energy-dependent Hamiltonian referred 
to the residual nucleus, belonging to the eigenvalues $E_R$ 
and $E_R+\omega$, respectively~\cite{bc,book}. The norm of 
$\phi^{}_{E_{\scriptscriptstyle R} n}$ is 1 and $S_n (E_R)$ is 
the spectroscopic factor associated with the removal process, 
i.e. it is the probability that the residual nucleus can 
indeed be conceived as the target nucleus with a hole. 
The dependence of 
$\chi^{\left( - \right)}_{p'_{\scriptscriptstyle N} 
E_{\scriptscriptstyle R} n}$ upon $p'_N$ is hidden in the asymptotic 
state $\vert \Psi^A_f \rangle$ and the boundary conditions 
are those of an incoming wave.

Because of the complexity of the eigenvalue problem in the
continuum, a complex mean-field interaction with energy-dependent
parameters is usually assumed between the residual nucleus 
and the emitted nucleon. Then,
$\chi^{\left( - \right)}_{p'_{\scriptscriptstyle N} 
E_{\scriptscriptstyle R} n} \sim 
\chi^{\left( - \right)}_{p'_{\scriptscriptstyle N}}$
and the nonlocality of the original Feshbach Hamiltonian is 
taken into account by multiplying the scattering wave by the 
appropriate Perey factor~\cite{perey}. Several models for this 
FSI are discussed in the literature 
(for a review, see Refs.~\cite{book,kelly}). Here, the eikonal approximation 
is adopted and is described in more detail in Sec.~\ref{sec:eikonal}. 
Finally, in Sec.~\ref{sec:tensor} we present the complete formula 
for the hadronic tensor used in the calculation.

\subsection{Current operator and spinor distortion}
\label{sec:current}

While new data for the (e,e$'$p) reaction have become available at 
very high proton energies~\cite{ne18a,ne18b,e89003}, it has also become 
evident that many ingredients of the theoretical calculations must 
be upgraded and made adequate for the new kinematical regime. In 
particular, a nonrelativistic reduction of the electromagnetic current 
operator is no longer reliable. Since all other ingredients entering the 
scattering amplitude will be deduced in a Schr\"odinger-like framework, we 
follow the approach of Ref.~\cite{kelly1}. 

It is well known that a four-component Dirac spinor $\Psi$, with positive- and 
negative-energy components $\psi_+$ and $\psi_-$, respectively, and satisfying 
a Dirac equation with energy eigenvalue $E$, mass $m$, scalar and vector potentials 
$S$ and $V$, respectively, can be written as
\begin{equation}
\Psi = \left ( \begin{array}{c} \psi_+ \\ \psi_- \end{array} \right) = 
\sqrt{\frac{E+m}{2m}} \  \left ( \begin{array}{c} 
{1\hspace{-2pt}\rule[0.08ex]{0.45pt}{1.5ex}\hspace{2pt}}
 \\ 
\displaystyle{\frac{\vec \sigma \cdot {\hat {\vec \pi}}}{E+m+S(r)-V(r)}} \end{array} 
\right ) \  D^{1/2}(r) \  \phi \  \equiv \  \Lambda ({\hat {\vec \pi}} , r) \  \phi 
\  , 
\label{eq:spinor}
\end{equation}
namely it can be represented
by the action of the operator $\Lambda ({\hat {\vec \pi}} , r)$ on the wave function 
$\phi$, that satisfies a Schr\"odinger-equivalent equation with central $U_C$ 
and spin-orbit $U_{LS}$ potentials, which can either be expressed in terms of $S$ 
and $V$ or replaced 
by intrinsically nonrelativistic potentials. The Darwin nonlocality factor 
\begin{equation}
D(r) = 1 + \frac{S-V}{E+m} 
\label{eq:nonloc}
\end{equation}
is related to $U_{LS}$ by 
\begin{equation}
U_{LS}(r) = - \frac{1}{2\mu} \frac{1}{r D} \frac{d D}{d r} \   , 
\label{eq:ls}
\end{equation}
where $\mu \sim m (A-1)/A$ is the reduced mass with $A$ the mass number. 

In Ref.~\cite{kelly1}, the calculations were performed in configuration space
by defining the effective current operator (omitting spin indices for
simplicity)
\begin{equation}
{\hat J}^{\mu}_{\rm eff} = \Lambda^{\dagger} (\vec p_N^{\, \prime}, r) 
\  \gamma^0 \  \Gamma^{\mu} \  \Lambda (\vec p_m, r) 
\label{eq:eff-kelly}
\end{equation}
and by evaluating the operator part $\vec \sigma \cdot {\hat {\vec \pi}}$ in the 
effective momentum approximation (EMA), i.e. by replacing the operator 
${\hat {\vec \pi}}$ with the 
momenta $\vec p_N^{\, \prime}, \vec p_m$ determined by asymptotic 
kinematics. After choosing one out of the three (on-shell) equivalent
expressions for the electromagnetic vertex function 
$\Gamma^{\mu}$~\cite{defor}, the
effective current operator ${\hat J}^{\mu}_{\rm eff}$ was reduced to a 
simple 2x2 matrix acting on the nucleon spins, using the standard 
representation for $\gamma$ matrices as 4x4 operators in terms of 2x2 Pauli 
spin matrices~\cite{bjorken}. 

Here, the scattering amplitude is worked out in momentum space. Therefore, the 
operator $\vec \sigma \cdot {\hat {\vec \pi}}$ becomes just a multiplicative factor 
with a 2x2 matrix structure in spin space acting on the nucleon spins. Consequently, 
Eq.~(\ref{eq:scattampl1}) can be specialized to a ``relativized 
Schr\"odinger framework'' by considering the following effective current
operator (omitting again spin indices for simplicity), 
\begin{equation}
{\hat J}_{\rm eff}^{\mu} (\vec p, \vec p^{\, \prime}, \vec q, \omega) = 
\frac{1}{2\pi^3} \  \int d \vec r \  e^{i (\vec p + \vec q - \vec p^{\,
\prime}) \cdot \vec r} \  \Lambda^{\dagger} (\vec p^{\, \prime}, r) 
\  \gamma^0 \  \Gamma^{\mu} \  \Lambda (\vec p, r) \   ,
\label{eq:eff-mine}
\end{equation}
as can be easily shown by starting from the expression of the scattering
amplitude in configuration space and applying the proper Fourier
transformations. The nonrelativistic limit of Eq.~(\ref{eq:eff-mine}) is
recovered by setting $E,E' \sim m$, and $S(r)=V(r)=0$. Inspection of 
Eqs.~(\ref{eq:nonloc}) and (\ref{eq:spinor}) indicates that in this limit 
the spinor-distortion operator no longer depends on $r$ and the 
Fourier transform in Eq.~(\ref{eq:eff-mine}) produces the well known 
$\delta (\vec p^{\, \prime} - \vec p - \vec q)$ accounting for momentum 
conservation~\cite{bgprep,book}. In the following, we will keep $D(r)=1$ for 
the scattering state, because the distortion of a high-energy ejectile 
will be approximated by a uniform damping in nuclear matter with $U_{LS}=0$ 
(see Sec.~\ref{sec:eikonal}).

The electromagnetic vertex function $\Gamma^{\mu}$ for an on-shell nucleon 
can be represented through three equivalent expressions related by the 
Gordon identity~\cite{defor}. Here, we choose the following
\begin{equation}
\Gamma^{\mu} = \gamma^{\mu} G_M (Q^2) - \frac{P^{\mu}}{2m} F_2(Q^2) \  ,
\label{eq:emvert}
\end{equation}
where $G_M$ is the nucleon magnetic form factor, $F_2$ is its Pauli form factor,
and $P^{\mu} = (E'+E, \vec p^{\, \prime} + \vec p)$. By inserting
Eq.~(\ref{eq:emvert}) in Eq.~(\ref{eq:eff-mine}), the scattering amplitude
becomes  
\begin{eqnarray}
 J^{\mu}_n (\omega ,\vec q, \vec p_N^{\, \prime}, E_R) &= &\int d \vec p 
\  d \vec p^{\, \prime} \  
\chi^{\left( -\right)\, *}_{p'_{\scriptscriptstyle N} 
E_{\scriptscriptstyle R} n} (\vec p^{\, \prime}) \  
{\hat J}_{\rm eff}^{\mu} (\vec p, \vec p^{\, \prime}, \vec q, \omega) \  
\phi^{}_{E_{\scriptscriptstyle R} n} (\vec p) \  [S_n(E_R)]^{1\over 2} \nonumber \\
&= &  \int d \vec p \  d \vec p^{\, \prime} \  
\chi^{\left( -\right)\, *}_{p'_{\scriptscriptstyle N} 
E_{\scriptscriptstyle R} n} (\vec p^{\, \prime}) \  
\sqrt{\frac{E+m}{2m}} \  \sqrt{\frac{E'+m}{2m}} \  
\frac{1}{(2\pi)^{\frac{3}{2}}} \nonumber \\
& &\Bigg\{ G_M(Q^2) \Bigg[ \delta_{\mu 0} \left( 
{\widehat D}^{\frac{1}{2}} \  + \  
\frac{\vec \sigma \cdot \vec p^{\, \prime}}{E'+m} 
\frac{\vec \sigma \cdot \vec p}{E+m} \, {\widehat D}^{-\frac{1}{2}} \right) + 
\delta_{\mu i} \left( \vec \sigma \,  
\frac{\vec \sigma \cdot \vec p}{E+m} \, {\widehat D}^{-\frac{1}{2}} \  + \  
\frac{\vec \sigma \cdot \vec p^{\, \prime}}{E'+m} \, \vec \sigma \, 
{\widehat D}^{\frac{1}{2}} \right) \Bigg] \nonumber \\
& &\  - \frac{P^{\mu}}{2m} F_2(Q^2) \Big[ {\widehat D}^{\frac{1}{2}} 
\  - \  \frac{\vec \sigma \cdot \vec p^{\, \prime}}{E'+m} 
\frac{\vec \sigma \cdot \vec p}{E+m} \, {\widehat D}^{-\frac{1}{2}} \Big] 
\Bigg\} \  
\phi^{}_{E_{\scriptscriptstyle R} n} (\vec p) \  [S_n(E_R)]^{1\over 2} \  ,
\label{eq:scattampl2}
\end{eqnarray}
where 
\begin{equation}
{\widehat D}^{\pm \frac{1}{2}} \equiv \frac{1}{(2\pi)^{\frac{3}{2}}} \int d
\vec r \  e^{i (\vec p + \vec q - \vec p^{\, \prime}) \cdot \vec r} \  
D^{\pm \frac{1}{2}} (r) 
\label{eq:ftd}
\end{equation}
are functions of $|\vec p + \vec q - \vec p^{\, \prime}|$. 

The nucleon form factors are taken from Ref.~\cite{MMD}, while the Coulomb
gauge is adopted to restore current conservation at the one-body level by
modifying the longitudinal component accordingly.

\subsection{The eikonal approximation}
\label{sec:eikonal}

Similar to the case of current operators, the high proton energies, that can
be reached in (e,e$'$p) reactions at the new experimental facilities, also demand a
suitable approach to the treatment of the proton scattering wave. 
Traditionally, the assumed mean-field interaction between the ejectile and the
residual nucleus has been described by complex spin-dependent optical potentials
with energy-dependent parameters constrained by fitting phase shifts and analyzing
powers of elastic (inelastic) (p,p) scatterings on the corresponding residual
nucleus. A Schr\"odinger equation with incoming wave boundary conditions for 
each partial wave of $\chi^{\left( - \right)}_{p'_{\scriptscriptstyle N}}$ is 
solved up to a maximum angular momentum $L_{\rm max}(p'_N)$ satisfying a
convergency criterion. Typically, this method has been successfully applied to
(e,e$'$p) reactions at proton momenta below 0.5 GeV/$c$ and 
$L_{\rm max}<50$~\cite{book,kelly}. 

At higher energies, the optical analysis of proton elastic scattering is
improved by the relativistic description via Dirac phenomenology. The scattering
wave is still expanded in partial waves, but each component solves the 
Dirac equation containing the scalar and vector Dirac
potentials~\cite{ohio}.

An alternative, simpler but powerful, method by 
Glauber~\cite{glauber} suggests that, when the proton is highly energetic,
the Schr\"odinger equation is reduced to a first-order differential equation
along the propagation axis $\hat z$, 
\begin{equation}
\left( \frac{\partial}{\partial z} - i p'_N \right) \chi = \frac{i}{2p'_N} U
\  \chi \   ,
\label{eq:glauber}
\end{equation}
with boundary conditions such that asymptotically $\chi \rightarrow 1$, i.e.
corresponding to an incoming unitary flux of plane waves. In the pure Glauber
model, $U(r)$ is determined in a parameter-free way starting from the elementary
free proton-nucleon scattering amplitude at the considered energy and then
averaging over all possible configurations of the spectator nucleons. 
For $p'_N \gtrsim 1$ GeV/$c$, the scattering amplitude is dominated by inelastic 
processes and $U(r)$ is supposed to be mostly sensitive to its 
imaginary part describing the absorption~\cite{le-le,br1,br2}. Moreover, the 
Glauber model predicts that the ratio between the real and imaginary parts of 
$U(r)$ equals the ratio between the real and imaginary parts of the average 
proton-nucleon forward scattering amplitude, which is expected to be anyway 
small beyond the inelastic threshold~\cite{le-le}. Therefore, we can safely 
assume $U(r) \sim i W(r)$. Then, the solution to Eq.~(\ref{eq:glauber}) looks 
like~\cite{br2}
\begin{eqnarray}
\chi^{\left( - \right)}_{p'_{\scriptscriptstyle N}} (\vec r) &= &
e^{\left( i \, \vec p_N^{\, \prime} \cdot \vec r + 
\textstyle{\frac{i}{2p'_N}} \int_z^{\infty} U(\vec r_\perp, z') \  d z' 
\right)} = e^{\left( i \, \vec p_N^{\, \prime} \cdot \vec r - 
\textstyle{\frac{1}{2p'_N}} \int_z^{\infty} W(\vec r_\perp, z') \  d z' 
\right)} \nonumber \\
&\equiv &e^{ \left( i \, \vec p_N^{\, \prime} \cdot \vec r \right)} \  
e^{ \left( -\vec p_{\rm I} \cdot \vec r \right)} \  ,
\label{eq:eikchi}
\end{eqnarray}
i.e. as a plane wave with a damping factor related to the absorption part of 
the residual interaction. 

The reliability of this eikonal approximation (EA), that has a long tradition of
successful results in the field of high-energy proton-nucleus elastic
scattering~\cite{glaurep}, has been tested in the context of knockout 
reactions and in the momentum range of interest here 
($0.6 \lesssim q \lesssim 1$ GeV/$c$) 
against solutions of the Schr\"odinger equation with nonrelativistic complex 
optical potentials up to $L_{\rm max} = 120$ partial waves~\cite{br3} (see 
also Ref.~\cite{jan1}). For 
increasing energies, the EA is supposed to become more and more reliable, 
despite the actual semirelativistic nature of the approach~\cite{glauber}. 
Moreover, for emitted protons with outgoing energy beyond the inelastic 
threshold and initially bound momentum below the Fermi surface 
($p_m \lesssim p_{\rm Fermi} \ll p'_N,q$, with $p_{\rm Fermi}$ the target Fermi
momentum: the same kinematic conditions of the E89003 experiment at 
CEBAF~\cite{e89003}), it has been shown that the proton angular distribution 
can actually be reproduced by representing the scattering wave as a plane wave 
with an additional damping~\cite{br1,br2}. After all, for a fast moving 
object the nuclear
density can be considered roughly constant (but for a small portion on the
surface) and the eikonal wave of Eq.~(\ref{eq:eikchi}) simply corresponds to the
solution of a Schr\"odinger equation inside homogeneous nuclear matter. In the
next Sec.~\ref{sec:damping}, a microscopic justification of the damping factor
will be given by a detailed description of the link between $p_{\rm I}$ and the
imaginary part of the self-energy of a nucleon moving inside nuclear matter.
Here, it is sufficient to say that for sake of simplicity the damping vector
will be kept parallel to the wave vector of the scattered particle, i.e. $\vec
p_{\rm I} \parallel \vec p_N^{\, \prime}$. 

The EA of Eq.~(\ref{eq:eikchi}) can be also formulated by saying that the
scattering wave is approximated by a plane wave with a complex momentum $\vec
p_f = \vec p_N^{\, \prime} + i \vec p_{\rm I}$ and normalized as
\begin{equation}
\chi^{\left( - \right)}_{p'_{\scriptscriptstyle N}} (\vec r) = 
e^{\left( -\vec p_{\rm I} \cdot \vec R \right)} \  
e^{ \left( i \, \vec p_f \cdot \vec r \right)} = 
e^{\left( -\vec p_{\rm I} \cdot \vec R \right)} \  
e^{\left( i \, \vec p_N^{\, \prime} \cdot \vec r - \vec p_{\rm I} \cdot 
\vec r \right)} \   ,
\label{eq:cpw-r}
\end{equation}
with $\vec R$ a constant vector with modulus equal to the nuclear radius. In
fact, for a propagation along the $\hat z$ axis, the wave enters the nucleus at
$\vec r = - \vec R \equiv (0,0,-R)$ with unitary modulus and leaves it at $\vec
r = \vec R \equiv (0,0,R)$ damped by $e^{-2\vec p_{\rm I} \cdot \vec R}$. In
order to consider the Fourier transform of Eq.~(\ref{eq:cpw-r}), an extended 
definition of the distribution $\delta$ of a complex variable is required. 
In the Appendix of Ref.~\cite{br3}, it is actually shown that such an extension 
is possible so that we can define the EA of the scattering wave in momentum 
space as
\begin{equation}
\chi^{\left( - \right)}_{p'_{\scriptscriptstyle N}} (\vec p^{\, \prime}) 
\equiv e^{-\vec p_{\rm I} \cdot \vec R} \  
\delta (\vec p_f - \vec p^{\, \prime})  \  ,
\label{eq:cpw-k}
\end{equation}
where now $\vec p^{\, \prime}$ is a complex vector. The extension of the matrix
element of Eq.~(\ref{eq:scattampl2}) to the complex plane in 
$\vec p^{\, \prime}$ is possible if the rest of the integrand is an analytic 
function asymptotically vanishing for 
$|\vec p^{\, \prime}| \rightarrow \infty$~\cite{br3}. It is rather easy to 
check that, apart from the $\delta$ distribution, the integrand of 
Eq.~(\ref{eq:scattampl2}) meets these requirements. Therefore, the scattering
amplitude in the EA (also with spin indices explicitly indicated) becomes
\begin{eqnarray}
(J^{\mu})^{}_{s'_{\scriptscriptstyle N} n} 
(\omega ,\vec q, \vec p_N^{\, \prime}, E_R) &\sim &e^{-\vec p_{\rm I} \cdot \vec R} \  
\int d \vec p \  \langle \, s'_N \vert {\hat J}^{\mu}_{\rm eff} 
(\vec p, \vec p_f, \vec q, \omega) \vert s_n \, \rangle  
\  \phi^{}_{E_{\scriptscriptstyle R} n} (\vec p) \  [S_n(E_R)]^{1\over 2} 
\nonumber \\
&= &\frac{e^{-\vec p_{\rm I} \cdot \vec R}}{(2\pi)^{\frac{3}{2}}} \  
\sqrt{\frac{E_f+m}{2m}}\  \int d \vec p  \  \sqrt{\frac{E+m}{2m}} \  
\phi^{}_{E_{\scriptscriptstyle R} n} (\vec p) \  [S_n(E_R)]^{1\over 2} 
\nonumber \\
& &\qquad \Bigg\{ G_M(Q^2) \Bigg[ \, \delta_{\mu 0} \left( 
{\widehat D}^{\frac{1}{2}} \  
\delta_{s'_{\scriptscriptstyle N} s_{\scriptscriptstyle n}} \  + \  
\langle s'_N | \, \vec \sigma \cdot \vec p_f^{\, *} \  
\vec \sigma \cdot \vec p \, | s_n \rangle \  
\frac{{\widehat D}^{-\frac{1}{2}}}{(E_f+m)( E+m)} \right) + \nonumber \\
& &\qquad \qquad \qquad \quad \delta_{\mu i} \left( \langle s'_N | \, 
\vec \sigma \  \vec \sigma \cdot \vec p \, | s_n \rangle \  
\frac{{\widehat D}^{-\frac{1}{2}}}{E+m} + \langle s'_N | \, \vec \sigma \cdot 
\vec p_f^{\, *} \  \vec \sigma \, | s_n \rangle \  
\frac{{\widehat D}^{\frac{1}{2}}}{E_f+m} \right) \Bigg] \nonumber \\
& &\qquad \  - \frac{p_f^{\mu}+p^{\mu}}{2m} F_2(Q^2) \Bigg[ 
{\widehat D}^{\frac{1}{2}} \  
\delta_{s'_{\scriptscriptstyle N} s_{\scriptscriptstyle n}} \ - \  
\langle s'_N | \, \vec \sigma \cdot \vec p_f^{\, *} \  \vec \sigma \cdot \vec
p \, | s_n \rangle \  
\frac{{\widehat D}^{-\frac{1}{2}}}{(E_f+m)(E+m)} \Bigg] \Bigg\} \   ,
\label{eq:scattampl3}
\end{eqnarray}
where $p^{\mu} = (E,\vec p), \, p_f^{\mu} = (E_f,\vec p_f),$ with 
$E=\sqrt{|\vec p|^2+m^2}, \, E_f=\sqrt{|\vec p_f|^2+m^2}$, and $s'_N, s_n$ are 
the projections of the spins of the detected proton and of the residual hole
with collective quantum numbers $n$, respectively. The Fourier transform of the
Darwin nonlocality factor, in agreement with Eq.~(\ref{eq:ftd}), is function of
$|\vec p + \vec q - \vec p_f| = \sqrt{|\vec p + \vec q - \vec p_N^{\, \prime}|^2
+ p_{\rm I}^2}$.

\subsection{Hadronic tensor}
\label{sec:tensor}

After summing over the undetected final states with quantum numbers
$n$ of the residual nucleus, the hadron tensor 
$W_{\lambda,\lambda'}$ in momentum space becomes
\begin{eqnarray}
W^{}_{\lambda,\lambda'} &= &(-)^{\lambda + \lambda'} e^\mu_\lambda 
e^{\nu \, *}_{\lambda'} \  e^{-2 \vec p_{\rm I} \cdot \vec R} \  
\sum_n \int d \vec p \  d \vec k \  \langle \, s'_N \vert 
({\hat J}^{}_{\rm eff})^{}_{\mu} \, (\vec p, \vec p_f, \vec q,\omega) \  
\vert s_n \, \rangle \  \phi^{}_{E_{\scriptscriptstyle R} n} (\vec p) \  
\phi^*_{E_{\scriptscriptstyle R} n} (\vec k) \  S_n (E_R) \nonumber \\
& &\mbox{\hspace{5.5truecm}} \langle \, s_n \vert 
({\hat J}^{}_{\rm eff})^{\dagger}_{\nu} \, (\vec k, \vec p_f, \vec q, \omega)  
\vert s'_N \, \rangle \nonumber \\
&\equiv &(-)^{\lambda + \lambda'} e^\mu_\lambda e^{\nu \, *}_{\lambda'} \  
e^{-2 \vec p_{\rm I} \cdot \vec R} \  \int d \vec p \  d \vec k \  
\langle \, s'_N \vert 
({\hat J}^{}_{\rm eff})^{}_{\mu} \, (\vec p, \vec p_f, \vec q, \omega) \  
S (\vec p, \vec k; E_R) \  ({\hat J}^{}_{\rm eff})^{\dagger}_{\nu} \, 
(\vec k, \vec p_f, \vec q, \omega) \  \vert s'_N \, \rangle \  , 
\label{eq:hadtens1}
\end{eqnarray}
where
\begin{equation}
S (\vec p, \vec k; E_R) = \sum_n \  S_n (E_R) \  
\phi^{}_{E_{\scriptscriptstyle R} n} (\vec p) \  \vert s_n \, \rangle \langle \, s_n \vert \  
\phi^*_{E_{\scriptscriptstyle R} n} (\vec k) 
\label{eq:myspec}
\end{equation}
is the hole spectral function discussed in the next 
Sec.~\ref{sec:spectral}. The isospin indices have 
been omitted for simplicity and, as before, the summation over $n$ runs over 
the undetected final states of the residual nucleus that are present at 
a given excitation energy $E_R$. 

The hole spectral function can be conveniently expanded in partial 
waves in a sp basis as
\begin{equation}
S (\vec p, \vec k; E_R) = \sum_{lj} \  \sum_{m_l, m'_l} \  \sum_{m_s, m'_s} \  
\left( l \textstyle{1\over 2} m_l m_s \vert j m \right) \  
\left( l \textstyle{1\over 2} m'_l m'_s \vert j m \right) \  S_{lj} (p,k;E_R) \  
Y_{l m_l} (\hat p) \  \vert m_s \, \rangle \langle \, m'_s \vert \   Y^*_{l m'_l} (\hat k) \  .
\label{eq:sflj}
\end{equation}
This expansion should not be confused with the sum in Eq.~(\ref{eq:myspec}): 
each $lj$ term contributes to the hadronic tensor and can come either from a 
quasihole state or from above the Fermi surface, depending on the 
excitation energy. 

The angular integrations in Eq.~(\ref{eq:hadtens1}) can be easily performed 
by noting that the square root of the Darwin nonlocality factor, 
$D^{\pm \frac{1}{2}}(r)$ in Eq.~(\ref{eq:ftd}), is not far from 1 which 
would yield a $\delta (\vec p^{\, \prime} - \vec p - \vec q)$ in 
momentum space (see Fig.~\ref{fig:figIII-1} in the next Sec.~\ref{sec:dr}). 
Therefore, because of Eq.~(\ref{eq:cpw-k}), we impose the constraint that 
the vector $\vec p$ in Eq.~(\ref{eq:hadtens1}) lies in the same direction 
as $\vec p_f - \vec q$, i.e.
\begin{equation}
\vec p \sim p \  \frac{\vec p_f - \vec q}{\vert \vec p_f - \vec q\vert} = 
\frac{p}{\sqrt{(\vec p_N^{\, \prime} - \vec q)^2 + p_{\rm I}^2}} \  
\left( \vec p_f - \vec q \right) \  ,
\label{eq:pappr}
\end{equation}
and similarly for $\vec k$. This approximation is reliable for high values of 
the involved momenta, as is the case for the kinematics of Ref.~\cite{e89003}. 
It is then easy to get rid of the angular integrations in Eq.~(\ref{eq:hadtens1}) 
so that the hadronic tensor, with explicit spin quantum numbers, takes the form 
\begin{eqnarray}
(W^{}_{\lambda,\lambda'})^{}_{s'_N} &= &(-)^{\lambda + \lambda'} e^\mu_\lambda 
e^{\nu \, *}_{\lambda'} \  e^{-2 \vec p_{\rm I} \cdot \vec R} \  
\sum_{lj} \  \sum_{m_l, m'_l} \  \sum_{m_s, m'_s} \  
\left( l \textstyle{1\over 2} m_l m_s \vert j m \right) \  
\left( l \textstyle{1\over 2} m'_l m'_s \vert j m \right) \  
Y_{l m_l} (\widehat{p_f-q}) \  Y^*_{l m'_l} (\widehat{p_f-q}) \nonumber \\
& &\int_0^\infty d p p^2 \int_0^\infty d k k^2 \  \langle \, s'_N \vert 
({\hat J}^{}_{\rm eff})^{}_{\mu} \, (p, \vec p_f, \vec q, \omega) \vert m_s \, \rangle \  
S_{lj} (p,k; E_R) \  \langle \, m'_s \vert ({\hat J}^{}_{\rm eff})^{\dagger}_{\nu} \, 
(k, \vec p_f, \vec q, \omega) \vert s'_N \, \rangle \  ,
\label{eq:hadtens2}
\end{eqnarray}
where ${\hat J}^{\mu}_{\rm eff} (p, \vec p_f, \vec q, \omega)$ is defined by 
inserting the approximation~(\ref{eq:pappr}) into Eq.~(\ref{eq:scattampl3}). 

Since the missing energy of the reaction is defined as~\cite{frumou,book} 
\begin{equation}
E_m = \omega - T_{p'_{\scriptscriptstyle N}} - T_R \   ,
\label{eq:missing}
\end{equation}
where $T_{p'_{\scriptscriptstyle N}}$ is the kinetic energy 
of the detected nucleon and 
\begin{equation}
T_R = \left[ p_m^2 + (M_R + E_R)^2 \right]^{1/2} - M_R -E_R 
\label{eq:kinres}
\end{equation}
is the kinetic energy of the residual nucleus, the scattering 
amplitude $J^{\mu}_{s'_{\scriptscriptstyle N} n} (\omega ,\vec q, 
\vec p_N^{\, \prime}, E_R)$ of Eq.~(\ref{eq:scattampl3}) can be 
conveniently made to depend on $(\omega ,\vec q, \vec p_m, E_m)$. 
Therefore, the differential cross section (and other related 
observables) for a given kinematics $(\omega, \vec q)$ and a 
knockout proton corresponding to a missing energy $E_m$ will be plotted as a 
function of the missing momentum $\vec p_m$. Older experimental 
(e,e$'$p) data at low proton energy were usually collected in 
the form of the so-called reduced cross section~\cite{bgprep,book}
\begin{equation}
n(\vec p_m, E_m) \equiv \frac{d \sigma}{d \vec p_e^{\, \prime} 
d \vec p_N^{\, \prime}} \  \frac{1}{K \sigma_{eN}} \   ,
\label{eq:redxsect}
\end{equation}
where $K$ is a suitable kinematic factor and $\sigma_{eN}$ is the 
elementary (half off-shell) electron-nucleon cross section, in order 
to reduce the information contained in a five-fold differential 
cross section to a two-fold function of $\vec p_m$ and $E_m$. 
Whenever needed, theoretical results will also be presented as 
reduced cross sections using the CC1 prescription~\cite{defor} for 
$\sigma_{eN}$ and the corresponding extrapolation 
$\bar \omega = E'_N - \bar E$ (with $\bar E=\sqrt{p_m^2 + m^2}$) 
for the off-shell nucleon. The electron distortion is 
included in the EMA by replacing $\vec q$ with an effective 
$\vec q_{\rm eff}$~\cite{gp} for the acceleration by the 
Coulomb field (in the following, for sake of simplicity 
the $_{\rm eff}$ subscript will be omitted).

\section{Quasi-hole and quasi-particle properties}
\label{sec:spectral}

The calculation of the (e,e$'$p) cross section is most easily performed
by employing the quasihole wave function of Eq.~(\ref{eq:specampl}). 
By considering Eq.~(\ref{eq:sflj}) in a sp basis 
with orbital angular momentum $l$, total angular momentum $j$, and 
momentum $p$, we can relate the spectroscopic amplitude to the spectral 
function in the following way
\begin{equation}
S_{lj}(p,k;E)= \sum_n \left \langle \Psi_0^{\rm A} \mid
a^{\dagger}_{klj} \mid \Psi_n^{{\rm A} - 1}\right \rangle
\left \langle  \Psi_n^{{\rm A} - 1} \mid a_{plj} \mid
\Psi_0^{\rm A} \right \rangle
\delta(E-(E_0^{\rm A}-E_n^{{\rm A} - 1})) \  , 
\label{eq:specl}
\end{equation}
where $a_{plj}$($a^{\dagger}_{klj}$) denotes the 
removal (addition) operator for a nucleon. 
The spectral function $S_{lj}(p,k;E)$ can be obtained from
the imaginary part of the corresponding sp propagator
$g_{lj}(p,k;E)$. This Green's function solves the Dyson equation
\begin{equation}
g_{lj}(p,k;E)=g_{lj}^{(0)}(p,k;E)+\int dp_1\ p_1^2 \int dp_2\ p_2^2\
g_{lj}^{(0)}(p,p_1;E) \  \Delta \Sigma_{lj}(p_1,p_2;E) \  
g_{lj}(p_2,k;E) \  , 
\label{eq:dyson}
\end{equation}
where $g^{(0)}$ refers to a Hartree-Fock propagator and
$\Delta\Sigma_{lj}$ represents contributions to the real and
imaginary parts of the irreducible self-energy, which go beyond
the Hartree-Fock approximation of the nucleon self-energy used to
derive $g^{(0)}$ (see below).
A brief summary of the calculation of the self-energy and the solution
of the Dyson equation are included below.
More details can be found in Refs.~\cite{mpd,po95}.

\subsection{Quasi-hole properties}
\label{sec:quasihole}

The self-energy is constructed by a two-step approach employing the
boson-exchange potential B as defined by Machleidt in Ref.~\cite{mach}.
The treatment of short-range correlations is taken into account by solving
the Bethe-Goldstone equation.
In the first step this equation is solved in nuclear matter 
at a certain density with a reasonable choice for the starting energy.
Employing a vector bracket transformation~\cite{wocl},
the corresponding ``Hartree-Fock'' self-energy contribution in momentum
space is calculated for ${}^{16}{\rm O}$ using harmonic oscillator wave
functions for the occupied states with oscillator
length $\alpha = 1.72$ fm$^{-1}$.
Since this ``Hartree-Fock'' self-energy is obtained from nuclear matter,
corrections need to be applied to reinstate the properties of the
${}^{16}{\rm O}$ Fermi surface. In addition, higher-order
self-energy contributions are included.
This procedure involves the calculation of the imaginary part of the
self-energy for two-particle one-hole (2p1h) and two-hole one-particle (2h1p)
intermediate states reached by the G-matrix interaction and
calculated in ${}^{16}{\rm O}$.
The intermediate particle states correspond to
plane waves and must be orthogonalized
to the bound sp states~\cite{borr}.
Pure kinetic energies are assumed for these particle states.
While this assumption is not very realistic
for the description of the coupling to low-lying states, it is quite adequate
for the treatment of tensor and short-range correlations.
From these imaginary contributions to the self-energy one can obtain the
corresponding real parts by employing the appropriate dispersion relations.
Since the ``Hartree-Fock'' part was calculated in terms of a 
G-matrix, it already contains the 2p1h contribution mentioned above but
generated in nuclear matter. The corresponding real part of
the self-energy as calculated in nuclear matter is then subtracted to
eliminate the double counting terms.
This procedure is quite insensitive to the original choice of density
and starting energy for the nuclear matter G-matrix~\cite{mpd,po95}.
For the determination of the $p$-shell quasihole wave functions only the real
part of the self-energy is relevant.
Collecting all the contributions to this self-energy one has
\begin{eqnarray}
{\rm Re}\ \Sigma_{lj}(p,k;E) & = & \Sigma^{HF}_{lj}(p,k) + 
{\rm Re}\ \Sigma^{2p1h}_{lj}(p,k;E) -
{\rm Re}\ \Sigma^{c}_{lj}(p,k;E) +
{\rm Re}\ \Sigma^{2h1p}_{lj}(p,k;E) \nonumber \\
& = & \Sigma^{HF}_{lj}(p,k) + {\rm Re}\ \Delta \Sigma_{lj}(p,k;E) \  , 
\label{eq:selfe}
\end{eqnarray}
with obvious notation.
In the last line of Eq.~(\ref{eq:selfe}) we have included (the real part of)
$\Delta\Sigma_{lj}$ which was anticipated in Eq.~(\ref{eq:dyson}).
This self-energy yields a complete treatment of the effect of short-range and
tensor correlations for a finite nucleus~\cite{mudi}.
The resulting wave functions for $p$-shell nucleons also yield an excellent
description of the shape of the experimental (e,e$'$p) cross 
section~\cite{previous}.
The self-energy in Eq.~(\ref{eq:selfe}) does not include an adequate 
description of the coupling of the nucleon to low-lying collective excitations
which strongly influence the spectroscopic factors~\cite{geur,badi1,badi2}.
This deficiency is not important for the present paper since we are
addressing the question of the reliable extraction of spectroscopic
factors from (e,e$'$p) data.
It is therefore of great importance that the theoretical wave functions
generated by Eq.~(\ref{eq:selfe}) are of equivalent quality to the 
empirical ones used in the analysis of the data~\cite{nikhef}.
Indeed, when these wave functions are used to fit the data~\cite{previous}, they
yield spectroscopic factors that are essentially identical to the ones
from the empirical analysis.

The solution of the Dyson equation was previously obtained in a basis generated 
by enclosing the system in a spherical box~\cite{mpd,po95}.
For the present paper the Dyson equation has been solved directly in
momentum space by performing the discretization for the relevant eigenvalue
problem.
For discrete solutions like the quasihole states in ${}^{16}{\rm O}$,
the Dyson equation yields the following eigenvalue equation
\begin{equation}
\frac{p^2}{2m} \langle \Psi^{A-1}_n \vert a_{plj} \vert \Psi^A_0 \rangle
+ \int^{\infty}_0 dk\ k^2 {\rm Re}\ \Sigma_{lj}(p,k;E_n)
\langle \Psi^{A-1}_n \vert a_{klj} \vert \Psi^A_0 \rangle
= E_n \langle \Psi^{A-1}_n \vert a_{plj} \vert \Psi^A_0 \rangle \  .
\label{eq:eigv}
\end{equation}
Discretizing the integration in Eq.~(\ref{eq:eigv}) yields a
straightforward diagonalization problem.
The resulting quasihole wave functions for the $p_{1/2}$ and $p_{3/2}$
states are used for the analysis presented in Sec.~\ref{sec:results}.
We have checked that the present solution method yields identical
results as compared to those from the ``box method.'' 

\subsection{Darwin nonlocality factor}
\label{sec:dr}

As discussed in Sec.~\ref{sec:current}, the Darwin nonlocality factor
given by Eq.~(\ref{eq:nonloc}) is required for a proper treatment of the
current operator at high proton energies.
It is clear from Eq.~(\ref{eq:ls}) that this nonlocality factor
is related to the spin-orbit potential.
This relation can therefore be used to derive the nonlocality factor
from the nucleon self-energy discussed in Sec.~\ref{sec:quasihole}.
A complication in deriving this result is that the self-energies
constructed for ${}^{16}{\rm O}$ are inherently nonlocal in coordinate
space.
This many-body nonlocality is already present when Fock terms to the
self-energy are considered.
Additional contributions are generated when higher-order terms 
are included as in Eq.~(\ref{eq:selfe}).
The terminology here may be confusing so it is useful to point out that
the Darwin nonlocality factor refers to 
the nonrelativistic reduction of the Dirac equation which
yields a nonlocal term in the Schr\"odinger equation when starting from a local
Dirac equation.

\begin{figure}[h]
\begin{center}
\epsfig{file=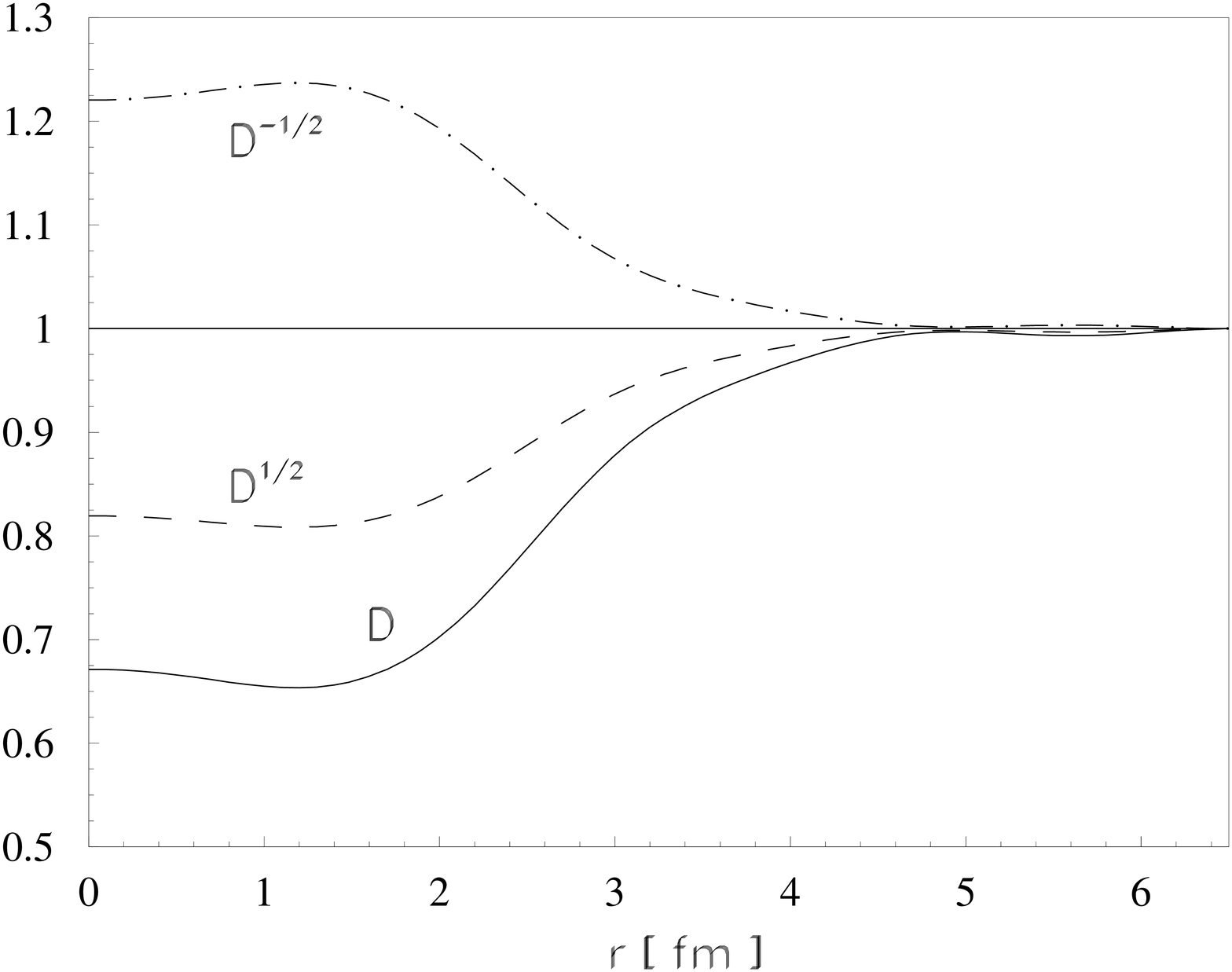, height=8.cm}
\end{center}
\caption{The Darwin nonlocality factor $D(r)$ obtained by inverting 
Eq.~(\ref{eq:ls}). Dashed and dot-dashed lines show $D^{\frac{1}{2}}(r)$ and 
$D^{-\frac{1}{2}}(r)$ entering Eq.~(\ref{eq:ftd}), respectively.}
\label{fig:figIII-1}
\end{figure}

Since Eq.~(\ref{eq:ls}) requires a local spin-orbit potential
we will construct local potentials from the nonlocal self-energies.
As shown in Ref.~\cite{borr}, it is possible to construct local 
potentials from the nonlocal self-energy by using the following
expression
\begin{equation}
{\rm Re}\ \Sigma^{local}_{lj}(r) = \int_0^{\infty} dr'\ r'^2\
{\rm Re}\ \Sigma_{lj}(r,r';E) \  ,
\label{eq:locpot}
\end{equation}
where the nonlocal self-energy in coordinate space is obtained from the
one in momentum space by a double Fourier-Bessel transformation given by
\begin{equation}
{\rm Re}\ \Sigma_{lj}(r,r';E) = \frac{2}{\pi}
\int_0^{\infty} dp\ p^2\ \int_0^{\infty} dp'\ p'^2\
j_l(pr)\ {\rm Re}\ \Sigma_{lj}(p,p';E)\  j_l(p'r') \  .
\label{eq:fbt}
\end{equation}
As shown in Ref.~\cite{borr}, a good representation to local potentials
given by Eq.~(\ref{eq:locpot}) is generated by a Woods-Saxon form.
Following this procedure we have obtained two local Woods-Saxon potentials
$V_{p1/2}(r)$ and $V_{p3/2}(r)$ for the relevant $p$-shell quasihole
states in ${}^{16}{\rm O}$, respectively.
Since the goal of the present work is to study the possibility to extract
spectroscopic factors at different $Q^2$, we have adjusted these 
potentials slightly to generate the correct experimental spin-orbit
splitting. In addition, we have ensured that the corresponding wave functions in
momentum space have the maximum overlap with those of the nonlocal
self-energies. 
These overlaps are given by 99.97\% for the $p_{1/2}$ and 99.99\% for
the $p_{3/2}$ wave functions, respectively.
It is now possible to decompose the local potentials
$V_{p1/2}$ and $V_{p3/2}$ in central and spin-orbit potentials
in the following way
\begin{equation}
V = U_0 + U_{LS} \vec L \cdot \vec S \  .
\label{eq:decomp}
\end{equation}
For $p$ states this implies that
\begin{equation}
U_{LS}(r) = \frac{2}{3} \left( V_{p\frac{3}{2}}(r) - V_{p\frac{1}{2}}(r) \right) \  .
\label{eq:uls}
\end{equation}
This potential can then be used to construct the Darwin nonlocality factor
by inverting Eq.~(\ref{eq:ls}). The latter is displayed in Fig.~\ref{fig:figIII-1} 
together with the functions $D^{\pm \frac{1}{2}}(r)$ used in Eq.~(\ref{eq:ftd}). The 
observed small deviation from unity of the latter functions is the basis for 
the approximation introduced in Eq.~(\ref{eq:pappr}) 
leading to the hadron tensor~(\ref{eq:hadtens2}).


\subsection{Damping of a quasi-particle}
\label{sec:damping}

As discussed in Sec.~\ref{sec:eikonal}, we will assume that the
damping of the nucleon on its way out of the nucleus is described
by a corresponding process taking place in nuclear matter.
Since the sp momentum is conserved in nuclear matter,
the propagation of a nucleon through nuclear matter is diagonal
in the sp momentum and can be represented by
\begin{equation}
g(p;E) = \int_{\epsilon_F}^{\infty}d\omega\ \frac{S_p(p;\omega)}
{E-\omega +i\eta}
+ \int_{-\infty}^{\epsilon_F}d\omega\ \frac{S_h(p;\omega)}
{E-\omega -i\eta} \  ,
\label{eq:gnm}
\end{equation}
where $S_p$ and $S_h$ (particle and hole spectral function)
describe the strength distribution above and below the Fermi energy
for a nucleon with sp momentum $p$.
These spectral functions have recently been determined self-consistently
by including the effects of short-range and tensor correlations in the
self-energy~\cite{libth}.
For this purpose the effective interaction is represented by the
equivalent of the T-matrix in the medium~\cite{di98,di99}.
The propagation of the nucleons determining this in-medium interaction
is also described by Eq.~(\ref{eq:gnm}) which includes full off-shell
effects.
The resulting interaction is employed to construct the nucleon
self-energy.
This self-energy is then used to solve the Dyson equation for nuclear
matter
\begin{equation}
g(p;E) = g^{(0)}(p;E) + g^{(0)}(p;E) \  \Sigma^{NM}(p;E) \  g(p;E) \  ,
\label{eq:dynm}
\end{equation}
where the unperturbed propagator is given by
\begin{equation}
g^{(0)}(p;E) = \frac{\theta (p-p_F)}{E-p^2/2m+i\eta}
+ \frac{\theta (p_F-p)}{E-p^2/2m-i\eta} \  .
\label{eq:gnon}
\end{equation}
The solution procedure for this problem involves several iteration steps
which are required because the solution to the Dyson equation already
appears in the determination of the effective interaction and
the self-energy, illustrating the nonlinearity of this problem.
The solution of Eq.~(\ref{eq:dynm}) can be written as
\begin{equation}
g(p;E) = \frac{1}{E-p^2/2m- \Sigma(p;E)} \  .
\label{eq:gsig}
\end{equation}
Taking advantage of the slow variation of the imaginary part of the
self-energy as a function of $p$, one can expand the self-energy at
the momentum $p_0$ for which
\begin{equation}
E \equiv \frac{p_0^2}{2m} + {\rm Re}\ \Sigma^{NM}(p_0;E) \  .
\label{eq:cpole}
\end{equation}
Performing the expansion in the square of this momentum and keeping
both the real and imaginary parts of the self-energy at $p_0$ plus
the first derivative of the real part, one obtains the so-called
complex pole approximation (CPA)~\cite{di98} for the propagator
which gives a very accurate representation of this quantity when
transformed to coordinate space.
For momenta above the Fermi momentum one has
\begin{equation}
g_{CPA}(p;E) = \frac{c_{p_0}}{p_0^2-p^2+i\gamma} \  ,
\label{eq:cpa}
\end{equation}
where
\begin{equation}
c_{p_0} = \left( \left. \frac{\partial {\rm Re} \Sigma^{NM} }
{\partial p^2} \right\vert_{p_0^2} \right)^{-1}
\label{eq:ce}
\end{equation}
and
\begin{equation}
\gamma = \left\vert {\rm Im} \, \Sigma^{NM}(p_0;E) \right\vert
\left( \left. \frac{\partial {\rm Re} \Sigma^{NM}}{\partial p^2}
\right\vert_{p_0^2} \right)^{-1} .
\label{eq:gam}
\end{equation}
This form of the propagator at a fixed energy has a simple pole
structure in the complex momentum plane.
The location of the relevant pole is given by
\begin{equation}
\kappa_0 = (p_0^4 + \gamma^2)_{}^{\frac{1}{4}} \  
e^{\frac{i}{2} \arctan \left( 
\frac{\gamma}{p_{\scriptscriptstyle{0}}^{\scriptscriptstyle{2}}} \right)} \  .
\label{eq:lpole}
\end{equation}


\begin{figure}[h]
\begin{center}
\epsfig{file=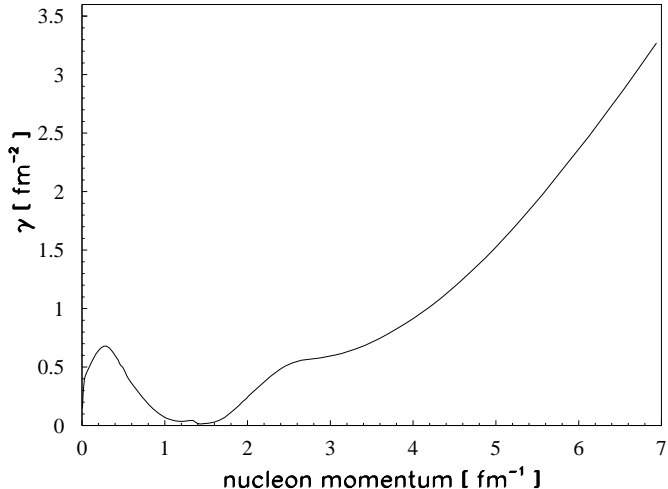, height=7.cm}
\end{center}
\caption{The quantity $\gamma$ of Eq.~(\ref{eq:gam}), related to the 
imaginary part of the nucleon self-energy, as function of 
$p_0 \sim p'_N$, the nucleon momentum.}
\label{fig:figIII-2}
\end{figure}


The imaginary part of this momentum is then used to describe the damping
in the eikonal approximation described in Sec.~\ref{sec:eikonal}, 
i.e. ${\rm Im} \, \kappa_0 \equiv p_{\rm I}$, with $p_{\rm I}$ from 
Eq.~(\ref{eq:eikchi}). 
To clarify the physics further one may
obtain the propagator in coordinate space using
\begin{eqnarray}
g_{CPA}(\vec r , \vec r^{\, \prime} ;E) & = & \frac{1}{(2\pi )^3}
\int d^3p\ e^{i \vec p \cdot ( \vec r - \vec r^{\, \prime})} g_{CPA}(p;E) 
\nonumber \\
& = & - \frac{c_{p_0}}{4\pi} \frac{e^{i \kappa_0 \vert \vec r - 
\vec r^{\, \prime} \vert}} {\vert \vec r - \vec r^{\, \prime} \vert} \  .
\label{eq:cpar}
\end{eqnarray}
From this result it is clear that the damping of the nucleon propagating
through nuclear matter is determined by the imaginary part of
$\kappa_0$, which in turn is determined by the imaginary part of the
self-energy at this energy. In Fig.~\ref{fig:figIII-2} the quantity $\gamma$ of 
Eq.~(\ref{eq:gam}), related to the imaginary part 
of the self-energy, is shown as a function of 
$p_0 \sim {\rm Re} \, \kappa_0 \equiv p'_N$. It is remarkable that at 
$p'_N \sim 7$ fm$^{-1}$, i.e. at the same proton kinematics of the NE18 
experiment~\cite{ne18a}, $\gamma$ is such that 
${\rm Im} \, \kappa_0 \equiv p_{\rm I} \sim 50$ MeV/$c$ gives the proper 
damping necessary to describe the observed absorption. In fact, in the 
context of the pure Glauber approximation $(W \propto p'_N)$ one would 
expect a higher proportionality factor, thus overestimating the 
quenching due to FSI (see Ref.~\cite{br2} and references therein). The 
outlined derivation of $\gamma$ gives a microscopic explanation for 
reducing this proportionality factor when embedding the travelling 
proton in nuclear matter.

This absorption effect of the medium is obtained for a realistic
nucleon-nucleon interaction~\cite{reid}.
Since this interaction is fitted to low-energy data, it is used
in the Lippmann-Schwinger equation to describe these data.
As a result, the coupling to intermediate states at higher energy
is constrained by the fit to these low-energy data.
Whether the description of these intermediate states as nonrelativistic
two-nucleon states is accurate is then less relevant.
One may also interpret the coupling to these intermediate 
(nonrelativistic two-nucleon) states as a
phenomenological way to include the coupling to inelastic channels,
quark effects, etc. 
For this reason we expect the present microscopic description of
nucleon absorption in the medium to be fully relevant for the
(e,e$'$p) reactions studied in this paper.

\section{Results}
\label{sec:results}

In this section we will discuss the results for the cross section of 
the $^{16}$O(e,e$'$p) reaction leading to the ground state and 
the first $\textstyle{3\over 2}^-$ excited state of the residual $^{15}{\rm N}$ nucleus. The 
main theoretical ingredient is the hadronic tensor of 
Eq.~(\ref{eq:hadtens2}), which describes the electromagnetic interaction 
assuming a relativistic one-body current operator including spinor 
distortion in the initial state only. This is consistent with the spin-orbit 
effects associated with the quasihole states in the residual nucleus 
[see Eq.~(\ref{eq:ls})]. The proton 
scattering wave is described in the eikonal approximation (EA), assuming 
a uniform and constant damping by nuclear matter through a 
nucleon self-energy containing the same short-range correlations used to 
generate the properties of the quasihole in the 
bound state. The electron wave is described through the EMA, that 
incorporates the acceleration due to the Coulomb field.


\begin{figure}[h]
\begin{center}
\epsfig{file=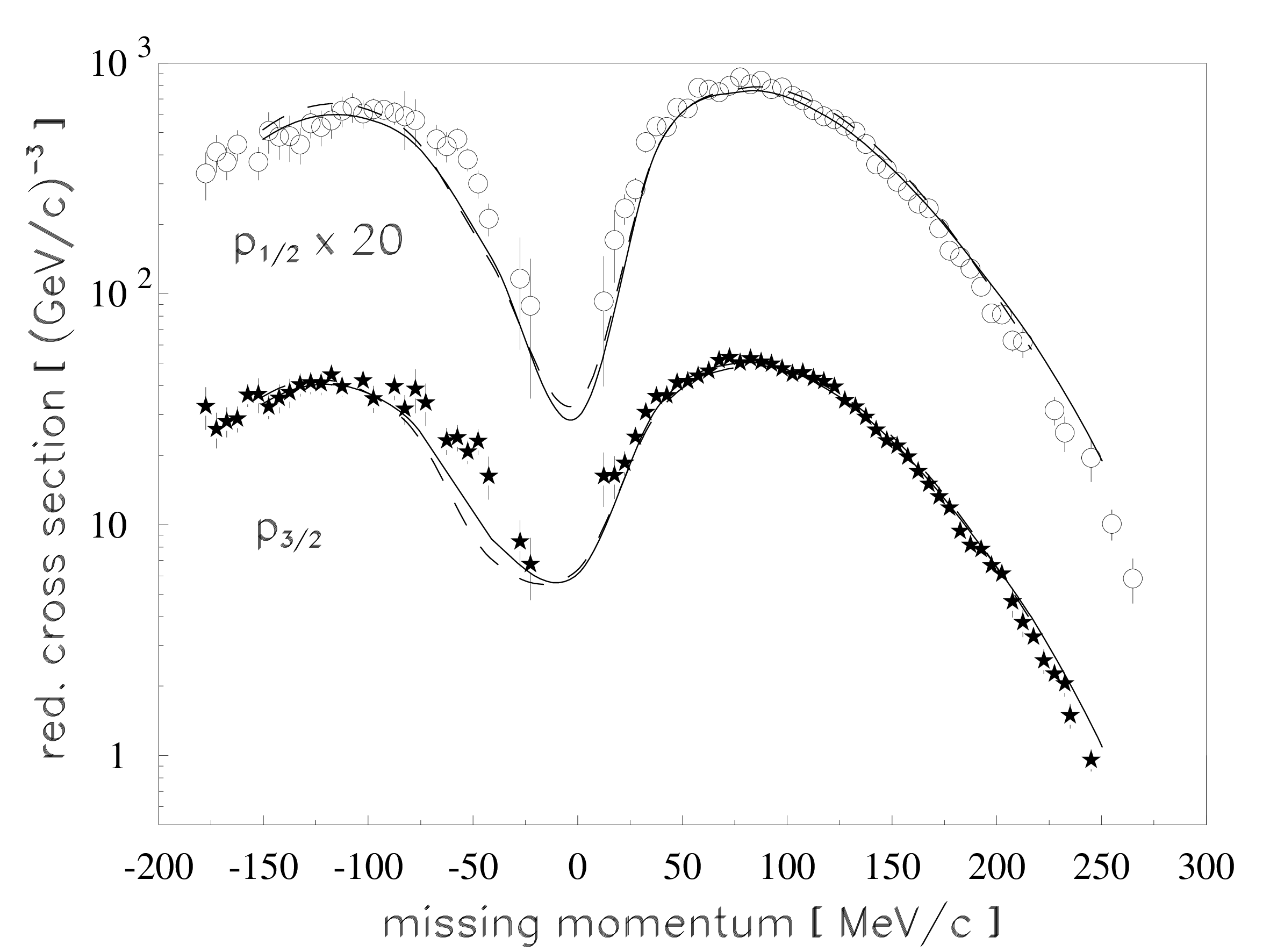, height=10.cm}
\end{center}
\caption{Cross section for the $^{16}{\rm O(e,e}'{\rm p)}^{15}$N 
reaction at $E_p = 90$ MeV constant proton energy in the center-of-mass 
system in parallel kinematics~[31]. Data for 
$p\textstyle{1\over 2}$ state have been multiplied by 20. Solid line is the result of 
Ref.~[30] using quasihole states; dashed line represents the same result but 
replacing the quasihole spectral function with the bound state from 
Ref.~[58]. All curves have been rescaled by the 
spectroscopic factors Z=0.644 and Z=0.537 for the $p\textstyle{1\over 2}$ 
and $p\textstyle{3\over 2}$ states, respectively.}
\label{fig:figIV-1}
\end{figure}


We first reconsider this reaction at low $Q^2$ using the conventional optical 
potential analysis for FSI~\cite{previous}. 
In Fig.~\ref{fig:figIV-1} the 
data from Ref.~\cite{nikhef} have been collected at a constant proton energy 
of 90 MeV in the center-of-mass system. They refer to the reduced cross 
section, defined by Eq.~(\ref{eq:redxsect}), as a function of the missing 
momentum $p_m$ in parallel kinematics, i.e. for 
$\vec p^{\, \prime}_N \parallel \vec q$. 
Therefore, the $p_m$ distribution can be obtained by increasing the momentum 
transfer $q$ from positive to negative values of $p_m$. Two transitions 
were considered, leading to the ground state $p\textstyle{1\over 2}$ and to the first excited 
state $p\textstyle{3\over 2}$ at $E_m=6.32$ MeV of $^{15}{\rm N}$. The data for the 
transition to the $p\textstyle{1\over 2}$ ground state have been multiplied by 20. The solid 
lines are the result of the calculation employing the quasihole part of the 
spectral function of Eq.~(\ref{eq:sflj}) for the $p\textstyle{1\over 2}$ and 
$p\textstyle{3\over 2}$ partial waves, respectively. The normalization of the curves 
is adjusted to fit the data indicating that the intrinsic normalization of the 
quasihole, 0.89 for the $p\textstyle{1\over 2}$ and 0.914 for the $p\textstyle{3\over 2}$, 
must be significantly 
reduced to $Z_{0p1/2}=0.644$ and $Z_{0p3/2}=0.537$, respectively, because only the 
depletion due to short-range correlations has been taken into 
account~\cite{previous}. Incidentally, long-range correlations spread the total 
$\textstyle{3\over 2}^-$ strength over three states in the discrete spectrum, 
so that the $p\textstyle{3\over 2}$ 
data account for 86\% of the strength only; by rescaling 
the spectroscopic factor by this fraction we get $Z_{0p3/2}=0.624$, in close 
agreement with the corresponding ground state value~\cite{previous}. The 
dashed lines in 
Fig.~\ref{fig:figIV-1} refer to the calculation including the same spectroscopic 
factors but replacing the quasihole bound state by the wave function of 
Ref.~\cite{Zhalov}, which is obtained by a Skyrme-Hartree-Fock method for 
$^{16}$O. Both descriptions are in 
very good agreement with the data, with a slight preference for the quasihole 
results at negative $p_m$.


\begin{figure}[h]
\begin{center}
\epsfig{file=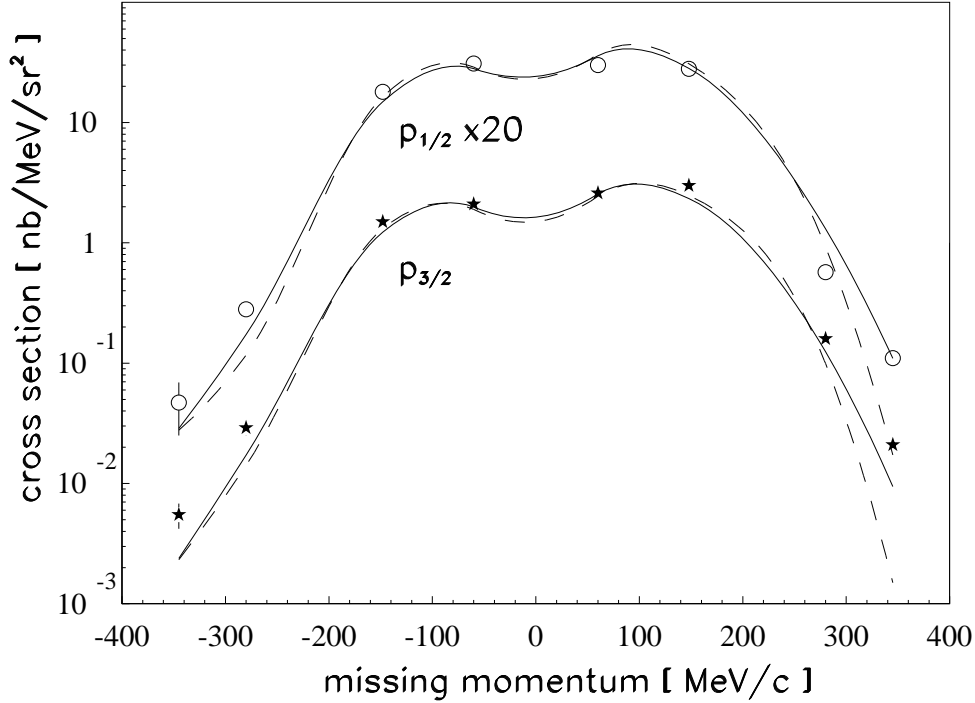, height=10.cm}
\end{center}
\caption{Cross section for the $^{16}$O(e,e$'$p)$^{15}$N 
reaction at $Q^2=0.8$ (GeV/$c)^2$ in perpendicular 
kinematics~[19]. Data for 
$p\textstyle{1\over 2}$ state have been multiplied by 20. The solid lines represent 
the result of the present calculation. The dashed lines are obtained by replacing 
the quasihole states with the bound state wave functions of Refs.~[58]. 
In all cases, the results have been 
rescaled by the same spectroscopic factors as in Fig.~3, 
namely Z=0.644 and Z=0.537 for the $p\textstyle{1\over 2}$ and 
$p\textstyle{3\over 2}$ states, respectively.}
\label{fig:figIV-2}
\end{figure}


In Fig.~\ref{fig:figIV-2} the same reaction is considered in a very different 
kinematical regime, namely at constant $(\vec q, \omega)$ with $Q^2=0.8$ 
(GeV$/c)^2$~\cite{e89003}. The data here refer to a five-fold differential 
cross section, avoiding any ambiguity in modelling the half off-shell elementary 
cross section $\sigma_{ep}$ of Eq.~(\ref{eq:redxsect}). Again, results for the 
transition to the ground state $p\textstyle{1\over 2}$ have been multiplied by 20. The 
theoretical calculations are displayed with the same notations as in 
Fig.~\ref{fig:figIV-1}, i.e. solid lines for the results with the quasihole 
bound state and dashed lines by employing the wave function of Ref.~\cite{Zhalov}. 
The preference for the first choice is here more evident. In any case, it is 
remarkable that the calculations reproduce the data by using the same 
spectroscopic factors as in the previous kinematics, i.e. $Z_{0p1/2}=0.644$ and 
$Z_{0p3/2}=0.537$. Therefore, contrary to the findings of Ref.~\cite{fsz}, 
we do not find any need for a $Q^2$ dependence of the spectroscopic factors over 
a wide kinematical range. This outcome is particularly welcome, since by 
definition these factors describe a spectroscopic nuclear property 
that must be independent of the probe scale $Q^2$. Finally, we conclude that 
the treatment of the bound state wave function is not responsible for the 
$Q^2$ dependence found in Ref.~\cite{fsz}.


\begin{figure}[h]
\begin{center}
\epsfig{file=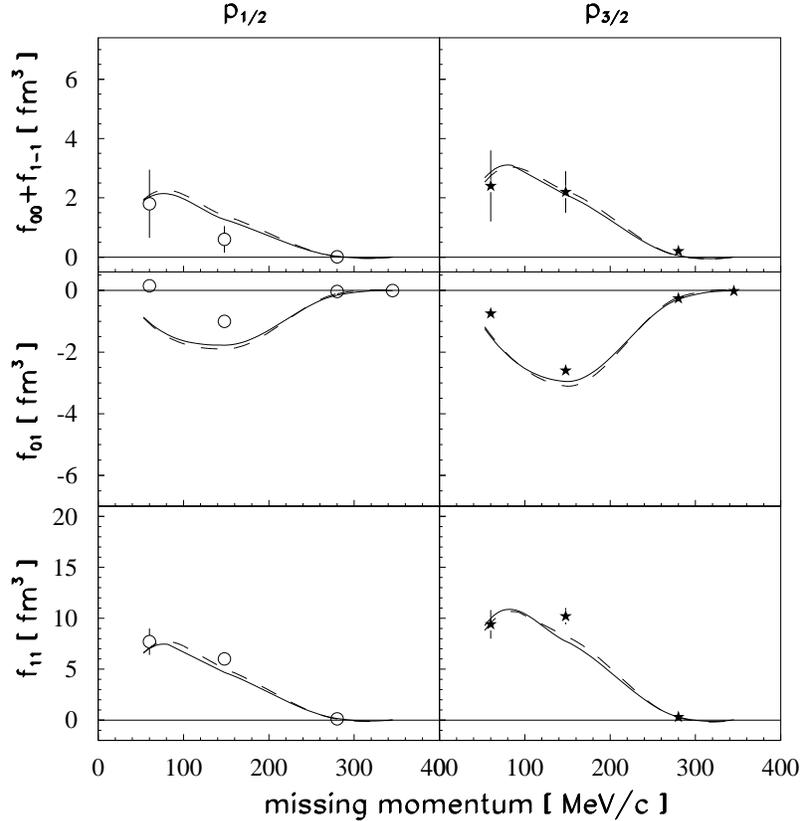, height=11.cm}
\end{center}
\caption{Structure functions for the $^{16}$O(e,e$'$p)$^{15}$N 
reaction at $Q^2=0.8$ (GeV/$c)^2$ in perpendicular 
kinematics~[19]. Same notations and scaling of curves as in 
Fig.~4.}
\label{fig:figIV-3}
\end{figure}


In Fig.~\ref{fig:figIV-3} we show the results for the structure functions 
$f_{\lambda, \lambda'}$ in the same kinematics and with the same notations as 
in Fig.~\ref{fig:figIV-2}. They are defined by 
\begin{eqnarray}
f_{00} &= &W'_{00} \nonumber \\
f_{11} &= &W'_{11} + W'_{-1-1} \nonumber \\
f_{01} &= &2 {\rm Re} \left [ W'_{01} - W'_{0-1} \right ] \nonumber \\
f_{1-1} &= &2 {\rm Re} \left [ W'_{1-1} \right ] \  ,
\label{eq:fmunu}
\end{eqnarray}
with $W'_{\lambda, \lambda'}$ the hadronic tensor in the proton center-of-mass 
system. It is related to $W_{\lambda, \lambda'}$ in the lab frame by the 
transformation $W_{\lambda, \lambda'} = e^{i \alpha (\lambda - \lambda')}\,  
W'_{\lambda, \lambda'}$, i.e. by a rotation around the $\vec q$ direction of the 
angle $\alpha$ between the lepton scattering plane and the plane formed by 
$\vec q$ and $\vec p^{\, \prime}_N$. The rotation affects only the interference components, 
so that the cross section~(\ref{eq:cross}) becomes~\cite{bgprep,book}
\begin{equation}
{ {d \sigma} \over {d \vec p_e^{\, \prime} d \vec p_N^{\, \prime} } }= 
{ e^4 \over {16 \pi^2}} {1 \over Q^4 p_e p'_e } \   \{  L_{00} f_{00} \  + 
\  L_{11} f_{11} \  + \  L_{01} f_{01} \cos \alpha \  + \  L_{1-1} 
f_{1-1} \cos 2\alpha \} \  ,
\label{eq:xfmunu}
\end{equation} 
i.e. it becomes parametrized in terms of the different components of the 
nuclear response $f_{\lambda, \lambda'}$ to the virtual photon probe in the 
spherical basis. The agreement with data is still good for both transitions over 
the whole $p_m$ range, except for $f_{01}$, the interference between the longitudinal 
and transverse responses, which is known to be particularly sensitive to 
relativistic effects~\cite{udiasb,kelly1,jan2}.


\begin{figure}[h]
\begin{center}
\epsfig{file=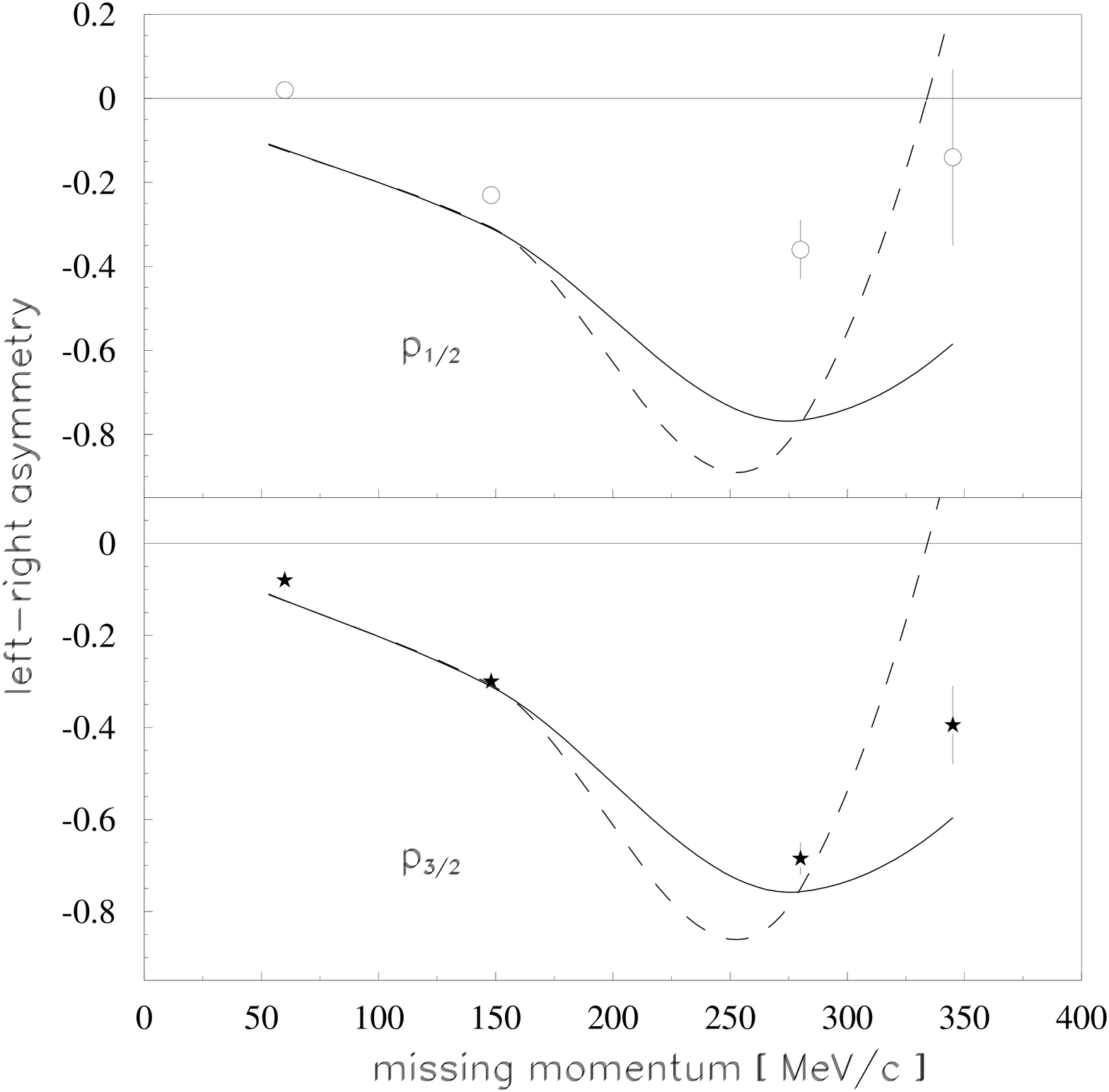, height=10.cm}
\end{center}
\caption{Left-right asymmetry for the $^{16}$O(e,e$'$p)$^{15}$N 
reaction at $Q^2=0.8$ (GeV/$c)^2$ in perpendicular 
kinematics~[19]. Same notations as in 
Fig.~4.}
\label{fig:figIV-4}
\end{figure}


Correspondingly, the left-right asymmetry
\begin{equation}
A_{LT} = \frac{ d\sigma (\alpha =0^0) - d\sigma (\alpha =180^0) }
              { d\sigma (\alpha =0^0) + d\sigma (\alpha =180^0) } 
\label{eq:asym}
\end{equation}
is displayed in Fig.~\ref{fig:figIV-4} in the same conditions and with 
the same notations as in Fig.~\ref{fig:figIV-2}. The discrepancy previously 
noted for $f_{01}$ is here amplified, particularly for the results with the bound 
state of Ref.~\cite{Zhalov}. A possible explanation is that only a full account 
of relativistic effects, specifically of spinor distortion in both bound and 
scattering states, is needed to reproduce the data~\cite{e89003,udiasb,kelly1}. 
In the present calculation this effect is included only for the 
bound state, while the Darwin nonlocality factor for the scattering state 
turns out to be 1 because the homogeneous damping in nuclear matter does not 
include spin-orbit contributions.

\section{Conclusions}
\label{sec:end}

We have developed a model for describing the (e,e$'$p) reaction at high $Q^2$ 
while linking it to nonrelativistic microscopic many-body ingredients like the 
quasihole spectral function. The goal is to critically consider the issue 
raised in Refs.~\cite{lapy,fsz} about a possible dependence of the spectroscopic 
factors upon $Q^2$. 

We use an unfactorized approach where, following Ref.~\cite{kelly1}, a 
relativistic one-body electromagnetic current operator is 
adopted in a Schr\"odinger-based framework avoiding any nonrelativistic reduction. 
The effect of spinor distortion by the Dirac scalar and vector potentials is 
consistently included only for the bound state by evaluating the Darwin nonlocality 
factor through the spin-orbit potential generated by the self-energy of the 
quasihole spectral functions. The proton scattering wave is described in an eikonal 
approximation (tested against DWIA solutions of a complex spin-dependent optical 
potential~\cite{br3}). The absorption is calculated by using a spectral function for 
nucleons in nuclear matter including the same short-range and tensor correlations 
adopted in the calculation of the nucleon self-energy in a finite 
volume for the $p$-shell quasihole states of $^{16}$O. 

In Ref.~\cite{previous} these bound state wave functions have been used to analyze 
the data for $^{16}$O(e,e$'$p) at low $Q^2$~\cite{nikhef} yielding a very good 
description of the reduced cross sections. In the present work, we have considered 
the recent data for the same reaction at higher $Q^2$~\cite{e89003} and we have 
performed the analysis using the same bound state wave functions and the same 
spectroscopic factors extracted from the low-$Q^2$ analysis. The description of the 
data at higher $Q^2$ is still very good regarding both the 5-fold differential 
cross section and the structure functions. Only the interference $f_{01}$ structure 
function, and the related left-right asymmetry $A_{LT}$, show a visible discrepancy, 
particularly for the $p\textstyle{1\over 2}$ state. A possible explanation could be related to our 
incomplete treatment of the relativistic effects because the spinor distortion of the 
final state is not considered. 

However, we emphasize that our consistent analysis of low- and high-$Q^2$ data using 
the same microscopic many-body ingredients for the quasihole states and the damping 
of the proton scattering wave allow us to conclude that we do not observe any $Q^2$ 
dependence of the spectroscopic factors over the considered wide range 
$0.02 \leq Q^2 \leq 0.8$ (GeV/$c$)$^2$. This outcome is most welcome, since by 
definition these factors describe a spectroscopic nuclear property that must be 
independent of the probe scale $Q^2$. Finally, since we get a very good description 
of the high-$Q^2$ data replacing our quasihole states with the bound states 
of Ref.~\cite{Zhalov}, we can also conclude that the quality of the wave 
functions is not responsible for the unexpected $Q^2$ dependence of the spectroscopic 
factors observed in Ref.~\cite{fsz}. 

\acknowledgments

This work is supported by the U.S. National Science Foundation under Grant No. 
PHY-9900713. We acknowledge the hospitality of the Laboratory of Theoretical 
Physics at University of Gent, where part of this work was done.

\end{document}